\documentclass[aip,
amsmath,
amssymb,
reprint
]{revtex4-1}

\usepackage{graphicx}
\usepackage{bm}
\usepackage{hyperref}

\usepackage[utf8]{inputenc}
\usepackage[T1]{fontenc}
\usepackage{mathptmx}

\usepackage[svgnames]{xcolor} 

\draft
\begin{document}

\preprint{AIP/123-QED}

\begin{minipage}{\textwidth}
\Large{The following article has been accepted by \textit{Chaos}. After it is published, it will be found at 
 \href{https://aip.scitation.org/journal/cha}{https://aip.scitation.org/journal/cha}, DOI: \href{http://dx.doi.org/10.1063/5.0012946}{10.1063/5.0012946}}
 \end{minipage}
 \vspace{1cm}
\title[Stability of a parametric harmonic oscillator with dichotomic noise]{Stability of a parametric harmonic oscillator with dichotomic noise}

\author{Daniel Schirdewahn}
\affiliation{ 
Institut für Physik und Astronomie, Universität Potsdam, Karl-Liebknecht-Stra\ss e 24-25, 14467 Potsdam, Germany. }%

\date{\today}

\begin{abstract}
The harmonic oscillator is a powerful model that can appear as a limit case when examining a nonlinear system.
A well known fact is, that without driving, the inclusion of a friction term makes the origin of the phase space -- which is a fixpoint of the system -- linearly stable. 
In this work we include a telegraph process as perturbation of the oscillator's frequency, for example to describe the motion of a particle with fluctuating charge gyrating in an external magnetic field. Increasing intensity of this colored noise is capable of changing the quality of the fixed point.
To characterize the stability of the system, we use a stability measure, that describes the growth of the displacement of the system's phase space position and express it in a closed form.  We expand the respective exponent for light friction and low noise intensity and compare both, the exact analytic solution and the expansion to numerical values. Our findings allow stability predictions for several physical systems. 
\end{abstract}

\maketitle

\begin{quotation}
The harmonic oscillator with stochastic influences is a common ``toy'' model for the description of complex physical systems, e.g. particles with varying charge gyrating in a planetary magnetic field. It has been shown, that a white noise perturbation of the frequency of a harmonic oscillator induces a bifurcation, changing the stability of the system's fixed point \cite{MallickMarcq2003}, while a colored noise term generally defies an exact analytic treatment \cite{MallickMarcq2004,MallickPeyneau2006}. We focus on the case of a dichotomic parametric noise, i.e. a frequency term switching between two states, and show, that it leads to the same bifurcation as the aforementioned parametric white noise. The stability of the system can be described by the growth rate of the distance of the system's current state from the phase space origin. We express this quantity in a closed analytic form. Our results can be used to predict the stability of several physical systems. 
\end{quotation}

\section{Introduction}
Studies of non-linear systems constitute a key-element of chaotic dynamics as an attempt to describe complex processes in nature and also society. A usual approach to serve this purpose is to investigate ``toy'' models which comprise and manifest typical properties of the underlying, more extensive systems.
Despite their simplicity, these models often elude their analytical solution, as shown by the three body problem, for instance. A further simplification is delivered by the linearization of the equations of motion in the vicinity of initial conditions, fixed points, periodic orbits or the like. The solutions now yield trends of the trajectory and allow to draw conclusions about the stability of the system. In this sense, the harmonic oscillator, naturally occurring by linearization, is one of the most powerful models in physics.

Another difficulty for the analysis of physical models are stochastic terms in the equations of motion.
These can be motivated by unknown deterministic influences or a high number of influencing variables, that justify a probabilistic description. These actions are often addressed by an additional stochastic force-like term, representing for example collisions with particles of the surrounding medium, that act as a heat bath. Still, there are cases were the random influences alter the system's potential, e.g. via changes in mass or charge, and lastly modify the frequency of the linearized harmonic oscillator. These influences appear as a multiplicative noise in the equation of motion. \\
Possible examples are, for instance:
\begin{itemize}
    \item {\it Astrophysics:} The gyrating motion of charged particles immersed in a magnetic field, as they occur as erosion-remains of Saturn's dense rings. These nano-grains, freed by micro-meteoroid impacts, may be trapped in wells of the effective potential \cite{Howard1999} and are subject to stochastic charging processes due to the planetary plasma environment \cite{Hsu2011, Hsu2018}. In such systems, questions arise regarding the time evolution of the particle trajectory and the stability of the system. As a first approximation, the potential wells may be assumed to be quadratics of the displacement from the equilibrium points.
    \item {\it Dynamical Systems:} Similar equations appear when looking for an estimation of the largest Lyapunov exponent of a high dimensional Hamiltonian system \cite{Casetti1995}. In that case, the displacement of the oscillator is a small perturbation of the system and the noise term describes the dynamics of the nonintegrable system. For a sufficiently high dimensional system with generic initial conditions, Gaussian white noise may be assumed (ib.). Still, one may find systems that require a non-continuous description of the Hamiltonian chaos. \\
    In this context, the growth rate of the oscillator is described by the Lyapunov exponent. Some publications keep that notion \cite{Zillmer2003,MallickMarcq2003}, as we do here as well.
    
\end{itemize} 

To date, studies, that examined multiplicative noise for the harmonic oscillator used  white and continuous colored noise \cite{MallickMarcq2003,MallickPeyneau2006}, while we aim to study noise terms that take discrete states, just as the charging of the nano-particles can no longer be approximated as continuous processes and has to be described as quantized steps. In this work, we focus on the limit case of noisy switching between two discrete states and find an analytic description for the stability of the system. A similar system has already been focused on in a  steady state setting, ensured by additional white noise forcing, that acted as a heat bath \cite{Bourret1973}. We omit such a stabilisation by an additional heating and use  a common stability measure -- the Lyapunov exponent -- to calculate the average growth rate of an energy-like quantity to characterize the long-term behavior of the parametric harmonic oscillator with a dichotomic noise.

\section{Characterization of the system}
We examine a harmonic oscillator with a multiplicative noise term that is described by the equation of motion
\begin{equation}
 \ddot x+\alpha\dot x+\big(1+\xi(t)\big)x=0\quad,\label{eq:dichotomicosci}
\end{equation}
given in a dimensionless form, where $x(t)$ is the displacement of the oscillator, $\alpha>0$ is a friction parameter, and $\xi(t)$ is a dichotomic stochastic process that switches between the states $\xi(t)=\xi_{1,2}$ with an average rate $\lambda_{1}>0$ from $\xi_1$ to $\xi_2$ and vice versa. \\
We make the simple assumption, that the transitions between states are independent events, which can be justified by turbulent environment and the molecular chaos of physical systems. To take account of this 'forgetfulness' of the system, the probability of an event in a small time interval $\delta t$ only depends on its length and is simply given as $\lambda_{1,2}\,\delta t$ (the index is depending on the current state of $\xi(t)$). \\
For the probability $\omega(\Delta t)$ of an event after a certain time span $\Delta t$, we divide the intervals into $N$ small sub-intervals and calculate the probability of only a single event in the $N$-th sub-interval. In the limit $N\rightarrow\infty$ we obtain an exponential distribution of the transition times $\Delta t_{1,2}$
\begin{equation}
  \omega_{\xi_{1,2}}(\Delta t)=\lambda_{\footnotesize 1,2}\mathrm{e}^{-\lambda_{\footnotesize 1,2} \Delta t}\quad.
\end{equation}
In the stationary limit, the probabilities of the states $\xi_{1,2}$ are given by the balanced rate equation 
\[
 0=-\lambda_1 p_{\xi_1}+\lambda_2 p_{\xi_2}
\]
with the normalisation $p_{\xi_1}+p_{\xi_2}=1$. They are proportional to the average duration of stay of the according state, which is just the inverse rate, $p_{\xi_{1,2}}\propto\overline{\Delta t_{1,2}}={\lambda_{1,2}}^{-1}$, normalised
\begin{equation}
 p_{\xi_1}=\frac{\lambda_{2}}{\lambda_1+\lambda_2}\quad,\quad\quad p_{\xi_2}=\frac{\lambda_{1}}{\lambda_1+\lambda_2}\quad.
\end{equation}
Without loss of generality, one can assume that $\lambda_1\geq\lambda_2$ and the averaged value of $\xi(t)$ vanishes\footnote{If the assumption of Eq. \ref{eq:zero-average} is not fulfilled, the random process can be transformed to $\xi(t)\rightarrow\frac{\xi(t)-\langle\xi(t)\rangle}{1+\langle\xi(t)\rangle}$, $\xi_{1,2}\rightarrow\frac{\xi_{1,2}-\langle\xi(t)\rangle}{1+\langle\xi(t)\rangle}$ and a renormalization of the time $t\rightarrow t\sqrt{1+\langle\xi(t)\rangle}$ leads again to the equation of motion (Eq. \ref{eq:dichotomicosci}) and $\langle\xi(t)\rangle=0$. This implies, that one value $\xi_{1,2}$ is negative, the other is positive.}:
\begin{equation}
 \langle\xi(t)\rangle=\frac{\xi_1\lambda_2+\xi_2\lambda_1}{\lambda_1+\lambda_2}=0\quad\label{eq:zero-average}
\end{equation}
These assumptions allow reducing the parameter set from four to three by defining $k=\lambda_2/\lambda_1\leq1$ and simultaneously $k=-\xi_2/\xi_1$. Now, one can describe the process $\xi(t)$ by $\lambda:=\lambda_1$, $\xi:=\xi_1$ and $k$. Notably, the limit case $k=0$ for a vanishing $\lambda_2$ \textit{is} the noiseless case, as at the same time $0=\xi_2=\langle\xi(t)\rangle$ (using Eq. \ref{eq:zero-average}).\\
For the probability $p_T(n)$ of  $n$ events in a fixed time interval $T$, we divide $T$ into sub-intervals and find the (binomial) distribution of $n$ events in the intervals, where we considered the probabilities of the system being in state $\xi_1$ or $\xi_2$. In the limit $N\rightarrow\infty$ we obtain a Poissonian
\begin{equation}
  p_T(n)=\mathrm{e}^{-\langle\lambda\rangle T}\frac{(\langle\lambda\rangle T)^n}{n!}
\end{equation}
with $\langle\lambda\rangle=p_1\lambda+p_2k\lambda=2\lambda/(1+k)$.
We use $p_T(n)$ to find the probability of an even or odd number of transitions in a time interval $\tau$ and average in the steady state limit to calculate the autocorrelation
\begin{equation}
 \langle\xi(t)\xi(t+\tau)\rangle=k\xi^2\mathrm{e}^{-\frac{2k}{1+k}\lambda\tau}\quad.
\end{equation}
In summary, $\xi(t)$ is a Poisson process, switching between its two states at exponentially distributed times. Its autocorrelation shows an exponential decay, thus it is a colored noise. The continuous process with the same autocorrelation behavior is the Ornstein-Uhlenbeck process, a stationary Gaussian process \cite{Gardiner2009}.

\section{The Lyapunov exponent of a parametric oscillator}
The energy of the normalised system is $E=\frac {x^2}{2}+\frac{\dot x^2}{2}$, which is proportional to the squared length of the phase space vector $\mathbf{X}=(x,\dot x)$. The Lyapunov exponent $\Lambda$, which is defined as 
\begin{equation}
 \Lambda=\lim\limits_{t\rightarrow\infty}\frac{1}{2t}\overline{\ln E}\quad,\label{eq:lyap_def}
\end{equation} 
can therefore be seen as an averaged growth rate of $\|\mathbf{X}\|$ and can describe the system's tendency to diverge from the origin of the phase space or to approach it, depending on its sign. In that way, we will use the Lyapunov exponent to measure the stability of the system.\\
A different way for calculating $\Lambda$ for the considered system is finding the ensemble average of 
\begin{equation}
 z=\frac{\dot x}{x}=\frac{\mathrm{d}}{\mathrm{d}t}(\ln x)
\end{equation}
in the stationary (or long time) limit \cite{MallickMarcq2003}: $\Lambda=\langle z\rangle=\int z P(z)\mathrm{d}z$ with the stationary probability density $P(z)$. One may illustrate that fact by pointing out, that Eq. \ref{eq:lyap_def} gives just the asymptotic behavior of $\frac{\mathrm{d}}{\mathrm{d}t}\ln\|\mathbf{X}\|$, while $\|\mathbf X\|$ grows with $x$. To prove this conjecture, knowledge about $P(z)$ is required, which will be calculated in the next section. The proof itself is shown in Appendix A.

\subsection{Solution of the Fokker-Planck equation}
To find the desired probability density $P(z)$, we first change variables from $z\in\mathbb{R}$ to $v\in[-1,1]$ to ease the numeric evaluation of occurring integrals. For that purpose, let $\phi\in[0,2\pi)$, defined by 
\begin{equation}
\cot\phi:=z\quad,                                   
\end{equation}
be a phase-like variable and $v=\cos\phi$. Then $v(z)$ has two branches $v=\pm z(1+z^2)^{-1/2}$, where the positive sign represents $\phi\in[0,\pi)$ and the negative sign represents $\phi\in[\pi,2\pi)$. In the following, we choose the second branch $v=-z(1+z^2)^{-1/2}$, as with this choice, $v$ increases with growing $z$ and the 'natural' sense of direction in the integrations will be preserved (see appendix B for more details). The equation of motion now translates to
\begin{eqnarray}
    \dot v=&\sqrt{1-v^2}-\alpha v\left(1-v^2\right)&+(1-v^2)^{3/2}\xi(t)\\
    =:&f(v)&+g(v)\xi(t).
    \end{eqnarray}
Note, that both, the It\={o} integral and Stratonovich integral, lead us to the same result due to the exponential autocorrelation: The additional terms cancelled out the diffusive terms, indicating, that in this context, the Kramers-Moyal expansion is not sufficient and a more elaborate method, illustrated by Horsthemke and Lefever\cite{Horsthemke2006} needs to be considered.\\
We define the joint probability $p_{i}(v),\,i=1,2$, of the processes $v(t)$ and $\xi(t)$ that is more exactly written as the transition probability
\[p_{i}(v,t):=p\big(v,\xi(t)\!=\!\xi_{i},\,t\,\big|\,v(0)\!=\!v_0,\,\xi(0)\!=\!\xi_0\big)\quad,\] where subscript $0$ indicates the initial values of $v(t)$ and $\xi(t)$, respectively. The associated Fokker-Planck equation reads \cite{Horsthemke2006,Anishchenko2003}
\begin{eqnarray*}
 \dot p_1(v)&=-\frac{\partial}{\partial v}\big(f(v)+g(v)\xi_1\big)p_1(v)-\lambda_1 p_1(v)+\lambda_2 p_2(v)\\
 \dot p_2(v)&=-\frac{\partial}{\partial v}\big(f(v)+g(v)\xi_2\big)p_2(v)+\lambda_1 p_1(v)-\lambda_2 p_2(v)\quad.
 \end{eqnarray*}
We introduce $P(v)=p_1(v)+p_2(v)$, which is the desired probability density, and $Q(v)=p_1(v)-p_2(v)$.  In the stationary limit, i.e. for vanishing time derivatives, the equations read
\begin{eqnarray}
  0&=&-\partial_v\left(P\frac{2f+g\xi(1-k)}{2}+Qg\xi\frac{1+k}{2}\right)\label{eq:deqPstat}\\
  0&=&-\partial_v\left(Q\frac{2f+g\xi(1-k)}{2}+Pg\xi\frac{1+k}{2}\right)\nonumber \\
  &&-P\lambda\left(1-k\right)-Q\lambda\left(1+k\right)\label{eq:deqQstat}\quad,
\end{eqnarray}
 where we use the reduced set of parameters $\xi$, $k$ and $\lambda$.
A direct integration of Eq. \ref{eq:deqPstat} yields the stationary probability flux
\begin{equation}
J=P\frac{2f+g\xi(1-k)}{2}+Qg\xi\frac{1+k}{2}=\mathrm{const.} 
\end{equation}
 that allows us to eliminate $Q$ from Eq. \ref{eq:deqQstat}. That way we end up with the final differential equation for $P$
\begin{eqnarray}
 \frac{\partial}{\partial v}\left[\frac{J\big(2f+g\xi (1-k)\big)-2P\big(f-g\xi k\big)\big(f+g\xi \big)}{g(1+k)}\right]\nonumber\\=-\frac{2\lambda}{g}(J-Pf)
\end{eqnarray}
and the solution
\begin{equation}
  P=J\frac{f+g\xi (1-k)}{(f+g\xi )(f-g\xi k)}+\frac{g(1+k)}{(f-g\xi k)(f+g\xi )}\Gamma_{\mathrm{part}}\quad,\label{eq:Psolution}
\end{equation}
with
 \begin{eqnarray*}
  \nonumber\Gamma_{\mathrm{part}}=&-J\exp\{\phi(v)\}[\int^v_{-1}\mathrm{d}x^{\prime\prime}\frac{\lambda g{\scriptstyle(\!x^{\prime\prime}\!)}\xi ^2k}{(f{\scriptstyle(\!x^{\prime\prime}\!)}-g{\scriptstyle(\!x^{\prime\prime}\!)}\xi k)(f{\scriptstyle(\!x^{\prime\prime}\!)}+g{\scriptstyle(\!x^{\prime\prime}\!)}\xi )}\\
  &\times\exp\{-\phi(x^{\prime\prime})\}+c]\\
  \phi(v)=&-\lambda(1+k)\int^v_{-1}\mathrm{d}x^{\prime}\frac{f{\scriptstyle(\!x^{\prime}\!)}}{(f{\scriptstyle(\!x^{\prime}\!)}-g{\scriptstyle(\!x^{\prime}\!)}\xi k)(f{\scriptstyle(\!x^{\prime}\!)}+g{\scriptstyle(\!x^{\prime}\!)}\xi )}\quad,
  \end{eqnarray*}
with a constant $c$. In the noise-less case $\xi(t)=\xi=0$, we expect the probability density to be $P(v)\propto\dot v^{-1}$, just as it would be for a usual harmonic oscillator. This condition fixes the constant to $c=0$ as in that case $\Gamma_{\mathrm{part}}\rightarrow0$ as $\xi\rightarrow0$ and $P(v)=\frac{J}{f(v)}\propto\dot v^{-1}$, as demanded.
Now the Lyapunov exponent  can be calculated by 
\begin{equation}
 \Lambda=\langle\frac{v}{\sqrt{1-v^2}}\rangle\quad\label{eq:LyapFormula},
\end{equation}
as $P_z(z)\mathrm{d}z=P_v(v)\mathrm{d}v$ and thus $\int z\,P_z(z)\mathrm{d}z=\int z(v)P_v(v)\mathrm{d}v$.

\section{Numerical evaluation}
\begin{figure}[htbp]
 \includegraphics[]{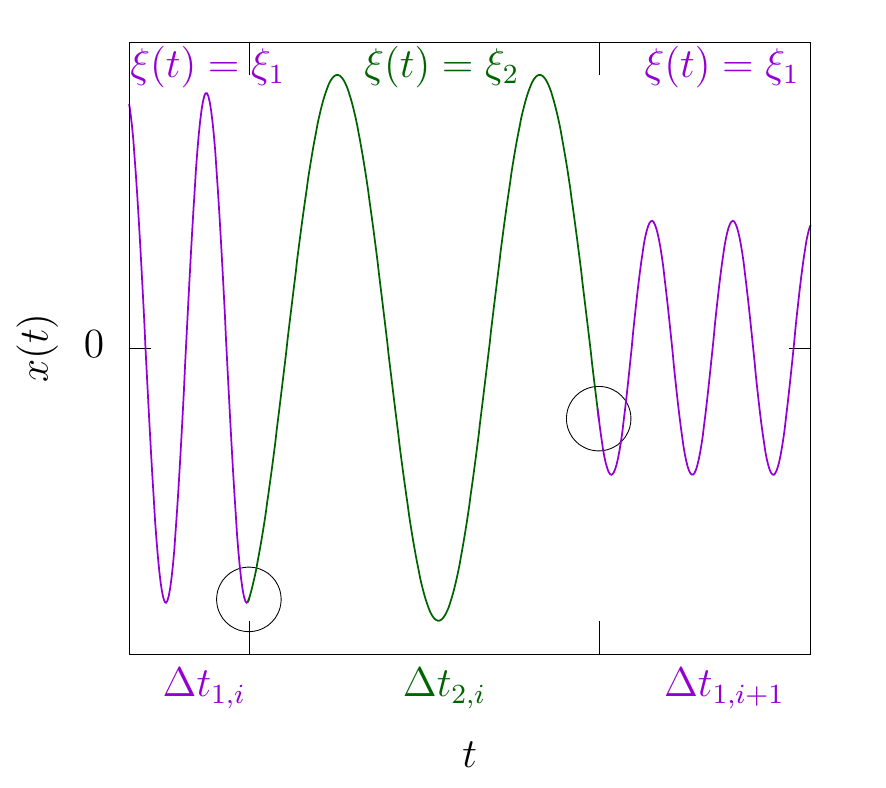}
\caption{Illustration of the numeric routine. During the (random) time intervals $\Delta t_{1,2}$ the trajectory $x(t)$ is a sinusoid with frequencies $\sqrt{1+\xi(t)}$, where $\xi(t)$ is either $\xi_{1}$ or $\xi_2$. At the transitions (marked by circles), the continuity of $x(t)$ and $\dot x(t)$ has to be ensured and determines the amplitude and phase of the oscillation.  
\label{fig:num_illus}}
\end{figure}

\begin{figure*}[hbtp]
\begin{minipage}[b]{\textwidth}
\begin{minipage}{0.02\textwidth}
 (a)\\\vfill
\end{minipage}\hspace*{1cm}
\begin{minipage}{.85\textwidth}
 \includegraphics[]{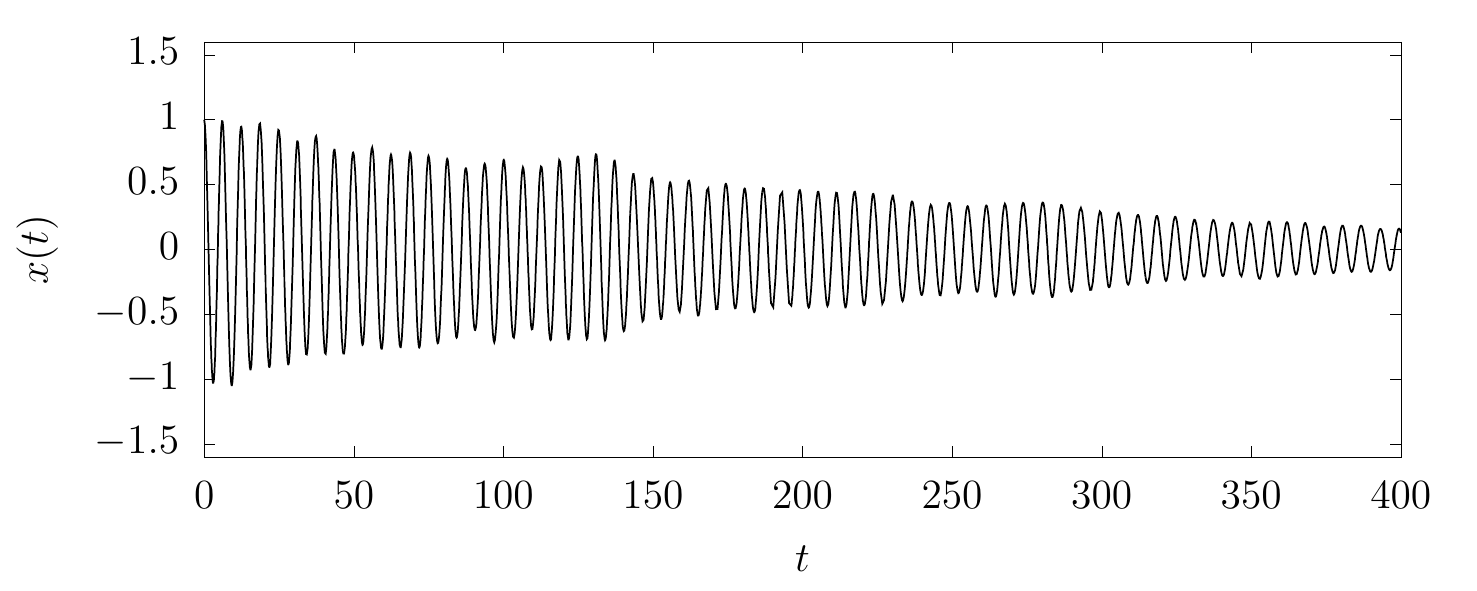}
\end{minipage}\\
\begin{minipage}{0.02\textwidth}
 (b)\\\vfill
\end{minipage}\hspace*{1cm}
\begin{minipage}[c]{.85\textwidth}
 \includegraphics[]{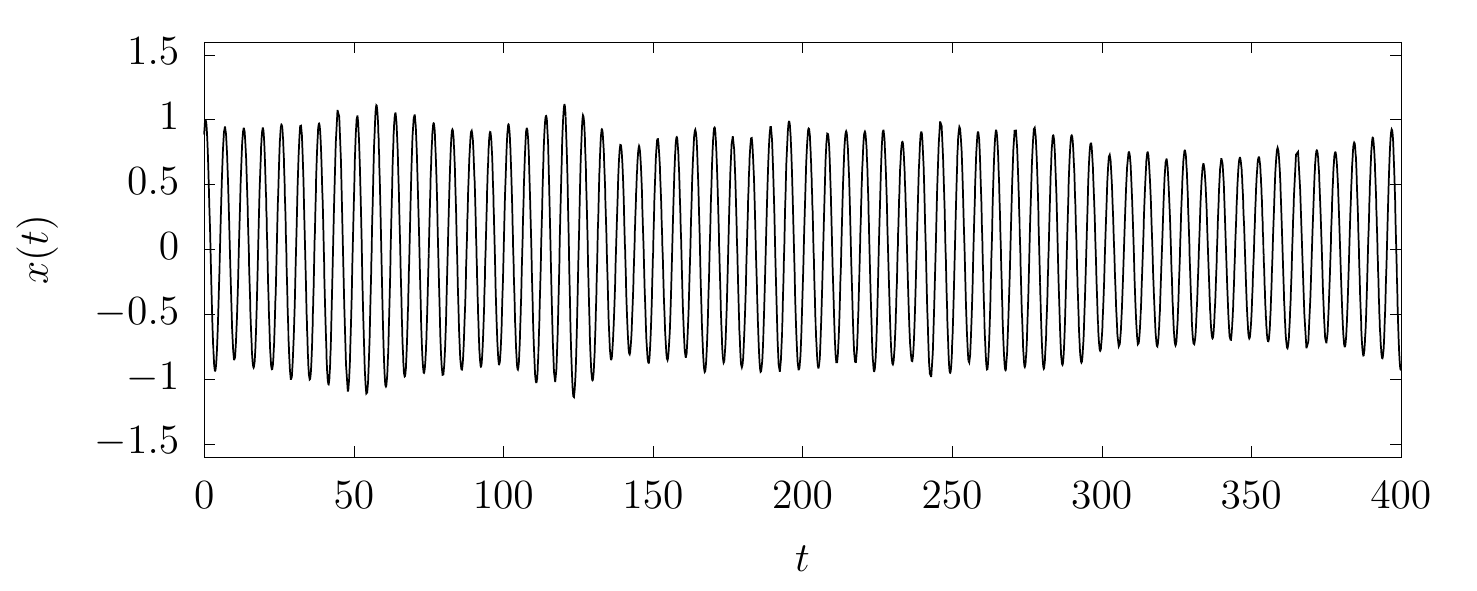}
\end{minipage}\\
\begin{minipage}{0.02\textwidth}
 (c)\\\vfill
\end{minipage}\hspace*{1cm}
\begin{minipage}[c]{.85\textwidth}
 {\includegraphics[]{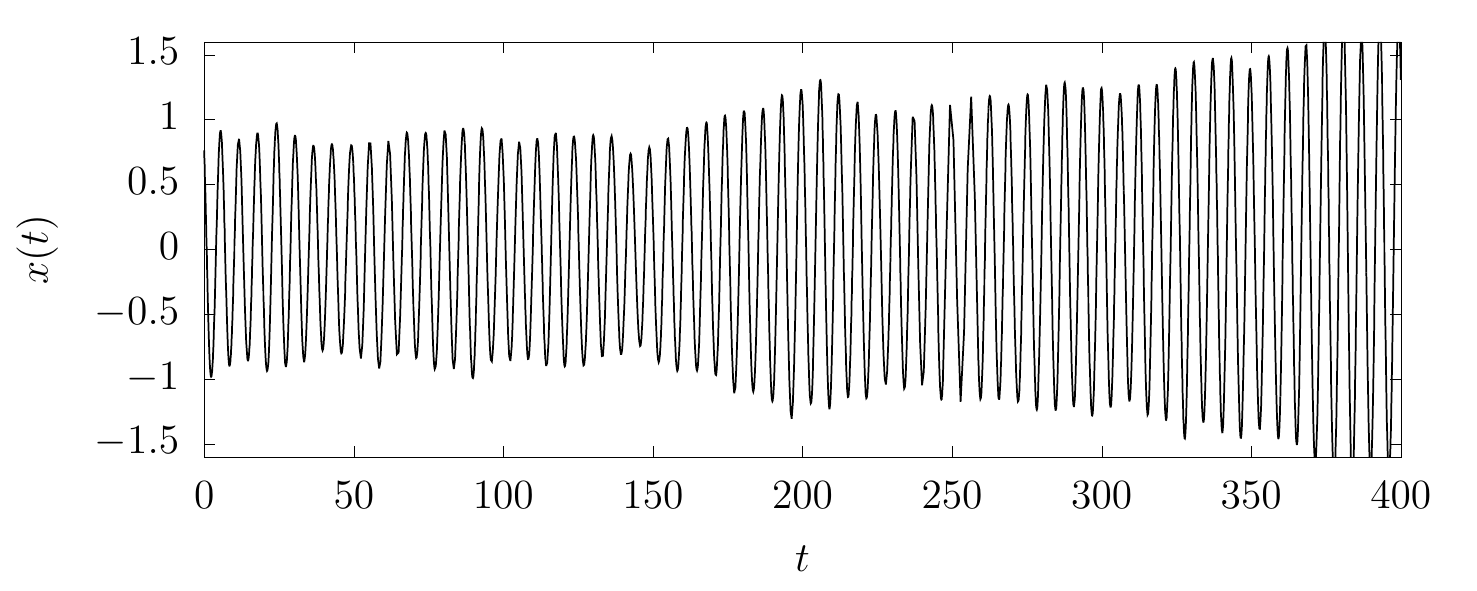}}
\end{minipage}
\end{minipage}
\caption{Exemplary trajectories $x(t)$ for fixed parameters $\xi=0.1$, $\lambda=1$, $k=1$ and different values of $\alpha$. (a) For a friction parameter $\alpha$ large enough, the amplitude decays due to friction losses and $x(t)$ approaches the resting point $x=0$. b) For a decreasing $\alpha$, $\Lambda$ is close to zero and the growth or decay rate of $x(t)$ is small. c) For small $\alpha$ the noise $\xi(t)$ pumps energy into the system that cannot be compensated by friction. The amplitude of $x(t)$ grows exponentially in the asymptotic limit.\\
The time $t$ is given in units of the  (averaged) period of the system, which is unity, as we considered a dimensionless equation of motion (Eq. \ref{eq:dichotomicosci}). As this differential equation is homogeneous and linear in $x$, there is no distinct length scale and $x(t)$ can be given in arbitrary units.\label{fig:num_traj}}
\end{figure*}

Realisations of the process $x(t)$ can be easily generated numerically, as each one is a piecewise compound of (known) solutions to the equation of motion (Eq. \ref{eq:dichotomicosci}) for $\xi(t)=\xi_1$ and $\xi_2$ with duration $\Delta t_1$ or $\Delta t_2$, respectively. These random transition times can be generated by library functions for exponential distributions and one only has to ensure the continuity of $x(t)$ and $\dot x(t)$ at the transition points. Fig. \ref{fig:num_illus} illustrates this ``concatenation'' of the solutions, while example trajectories are given in Fig. \ref{fig:num_traj}

The Lyapunov exponent can be found for example utilizing the definition in Eq. \ref{eq:lyap_def} by averaging the value $\frac{1}{2t_{\text{max}}}\ln E(t_{\text{max}})$ after a certain observation time $t_{\text{max}}$ for several realisations of the process $x(t)$
. We will compare the numerically determined Lyapunov exponent to the analytic solution in the next section, together with an expansion for small values of $\xi$ and $\alpha$.

\section{Small parameter expansion of the solution}
\begin{figure}[tbp]
 \includegraphics[]{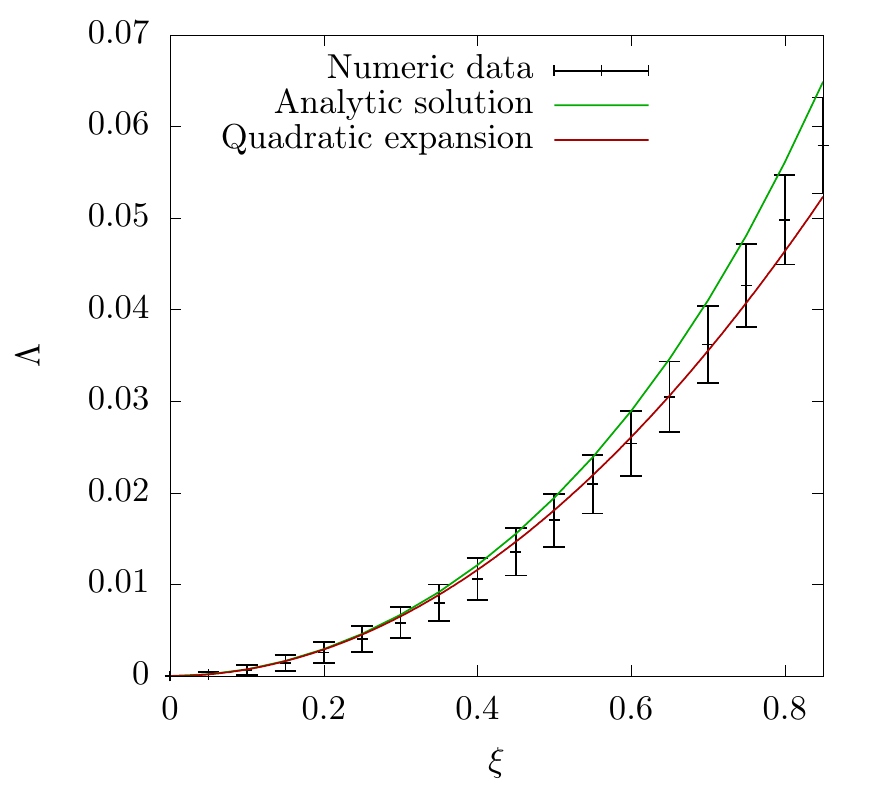}
\caption{Analytic solution of the Lyapunov exponent $\Lambda$ without friction (green line) and quadratic expansion with respect to the noise strength $\xi$ (red line), calculated according to Eq. \ref{eq:LyapFormula} and \ref{eq:gammaexpr}, respectively. The values for the asymmetry and rate parameters are $k=1$ and $\lambda=1$. The exact solution systematically overestimates the numerically determined value of $\Lambda$. The quadratic expansion does not follow this systematic behavior and falls below the numeric trend for $\xi>0.65$. In the evaluated range $0\leq\xi\leq0.85$, both analytic curves show a fair agreement and lie within a $15\%$ band of the numerical results. \label{fig:num_smallxi}}
\end{figure}

In the following we evaluate how $\Lambda$ behaves for a small friction parameter $\alpha$ and small noise strength $\xi$ and find a critical curve $\Lambda(\alpha,\xi)=0$, which separates the parameter space into a stable ($\Lambda<0$) and an unstable ($\Lambda>0$) domain.\\
As mentioned before $P_{\mathrm{\xi=0}}=\frac{J}{f(v)}$ and one finds easily, by linearizing in $\alpha$ and integrating, that
\begin{equation}
 \Lambda_{\mathrm{\xi=0}}=-\frac{\alpha}{2}+\mathcal{O}(\alpha^3)\quad,
\end{equation}
which is simply the relaxation rate of an undriven harmonic oscillator with the influence of friction.\\
As the first summand in Eq. \ref{eq:Psolution} has no term linear in $\xi$ and $\Gamma_{\text{part}}\propto \mathcal{O}(\xi^2)$ one can exclude mixed terms if expanding to second order and one only needs to find the coefficient $\gamma$ in the expansion
\begin{equation}
 \Lambda=-\frac{\alpha}{2}+\frac{\gamma}{2}\xi^2+\mathcal{O}(\alpha^n\xi^m)\quad,\,m+n\geq3\quad.
\end{equation}
Expanding Eq. \ref{eq:LyapFormula}, considering the $\xi$-dependency of $J$, we find
\begin{equation}
 \gamma=k\frac{(1+k)\lambda}{2\big(4+(1+k)^2\lambda^2\big)}+4k\frac{1-\mathrm{e}^{-\pi(1+k)\lambda}}{\pi\big(4+(1+k)^2\lambda^2\big)^2}\quad.
 \label{eq:gammaexpr}
\end{equation}
Thus, the critical curve $\Lambda(\alpha,\xi)=0$ is given by $\alpha_c=\gamma\xi_c^2$. For small values of the friction parameter, $\alpha<\alpha_c(\xi)$, $\Lambda>0$ and the system is unstable as its energy grows; the origin is an unstable fixpoint. For larger values $\alpha>\alpha_c(\xi)$, the energy decays and the origin is a stable fixpoint.
This transition behavior of $\Lambda(\alpha,\xi)$ resembles the linearized stochastic oscillator described by Mallick and Marcq \cite{MallickMarcq2003}, the small energy limit of a nonlinear oscillator with multiplicative white noise. In that case the relation of the critical friction coefficient $\alpha_c$ and the (quadratic) noise intensity $\Delta$ reads $\alpha_c=\Delta/4$ in the first order (Eq. 47 in \cite{MallickMarcq2003}). \\
Fig. \ref{fig:num_smallxi} shows the analytic solution of Eq. \ref{eq:LyapFormula} and the expansion $\Lambda=\frac{\gamma}2 \xi^2$ compared to numerically values, calculated from $10^5$ trajectories. The quadratic expansion needs significantly less computation time than the exact solution, as no integrals need to be evaluated. The exact solution systematically exceeds the numeric values, as the latter only converge towards the real value of $\Lambda$ for an infinite observation time $t_{\text{max}}\rightarrow\infty$. In the considered range $\xi\in[0;\,0.85]$ the quadratic expansion is even a better approximation to the numerical values and does not show a systematic deviation.\\
\section{Implications for nonlinear systems}
\label{sec:nonlin}
For the case of white parametric noise, it has been shown, that the inclusion of terms of higher order, namely $\propto x^3$ in the equation of motion Eq. \ref{eq:dichotomicosci}, energy dissipation can balance the pumping of the noise and a regime of stable, noisy oscillations may exist \cite{MallickMarcq2003}. This transition happens at exactly the moment, the origin of the linearized (harmonic) oscillator changes its stability.\\
In the comparable case of a deterministic chaotic system, a pair of complex conjugate eigenvalues of the Jacobian of the system cross the imaginary axis and a limit cycle can arise from a equilibrium of the system in a Poincar\'e-Andronov-Hopf bifurcation\cite{Lopez2015} (or shorter: Hopf bifurcation). This coins the term of a stochastic Hopf bifurcation for the systems like the nonlinear oscillator with parametric noise \cite{MallickMarcq2003}.
\section{Conclusion}
We considered the basic physical model of a harmonic oscillator with a frequency switching between two states at random times. We found the stationary probability density of an associated phase-like variable and used it to calculate the Lyapunov exponent of the system.\\
For a simple harmonic oscillator with friction, the origin of the phase space ($x=\dot x=0$) is a globally stable fixed point. With increasing multiplicative dichotomic noise, this feature disappears and after a certain critical value the origin is the only stable initial condition, while for all other values the energy rises and the phase-space coordinates will grow beyond any given limit.\\
In the future, we plan to use these results to characterize the stability of the trajectories of nanometer sized particles in the dust environment of Saturn's rings. In this setting, the  Lyapunov exponent will serve to give a time scale of the particles leaving their source region and in this way eroding the rings and limiting rings' lifetime.\\
For a more detailed description, we plan to include nonlinear terms, as described in section \ref{sec:nonlin}. It has to be evaluated, whether the energy injection from the dichotomic noise will be balanced by diffusive terms in the same manner as for white noise, leading to a comparable bifurcation.\\
The dichotomic noise described in the present article is a limit case to a Poissonian random walk. We contemplate to investigate this kind of noise in future works both analytically and numerically, where our presented results will serve as testing limit case.

\begin{acknowledgments}
The author would like to thank Prof. Frank Spahn for his supervision, support, and enriching discussions, as well as Dr. Holger Hoffmann and Prof. Ralph Metzler for constructive recommendations.\\
This work was written within the scope of a project supported by the Deutsche Forschungsgemeinschaft (Sp384/33-1).
\end{acknowledgments}

\section*{AIP Publishing Data Sharing Policy}
The data that support the findings of this study are available from the corresponding author upon reasonable request.

\appendix

\section{}

To show, that $\Lambda=\langle\cot\phi\rangle$, we find $\frac{\mathrm{d}}{\mathrm{d}t}\overline{\rho}$ for the radial variable $\rho=\frac12 \ln E$ and $\frac{\mathrm{d}}{\mathrm{d}t}\overline{\ln|\sin\phi|}$ by averaging $\rho$ and $\ln|\sin\phi|$ with respect to $\dot P(\rho,\phi)$, which is generally given by the r.h.s. of Eq. \ref{eq:deqPstat}.\\
After integration by parts and by using the boundary conditions $\lim\limits_{\rho\rightarrow\pm\infty}P=0$ and $P(\rho,0)=P(\rho,2\pi)$ we find 
\begin{widetext}
 \begin{eqnarray}
  \frac{\mathrm{d}}{\mathrm{d}t}\overline{\rho}=&\int\mathrm{d}\rho\mathrm{d}\phi\dot P\nonumber\\
  =&\int\mathrm{d}\rho\mathrm{d}\phi\left[-P\alpha\cos^2\phi+\left(\frac{\xi_++\xi_-}{2}P+\frac{\xi_+-\xi_-}{2}Q\right)\sin\phi\cos\phi\right]\nonumber\\
  \frac{\mathrm{d}}{\mathrm{d}t}\overline{\ln|\sin\phi|}=&\int\mathrm{d}\rho\mathrm{d}\phi\left[P\left(1+\alpha\sin\phi\cos\phi\right)\cot\phi-\left(\frac{\xi_++\xi_-}{2}P+\frac{\xi_+-\xi_-}{2}Q\right)\sin\phi\cos\phi\right]\nonumber\\
  =&\overline{\cot\phi}-\frac{\mathrm{d}}{\mathrm{d}t}\overline{\rho}\quad.
\end{eqnarray}
\end{widetext}
Note the dependencies inside the integrals $P=P(\rho,\phi)$ and $Q=Q(\rho,\phi)$, which are not written for brevity.
Again following the arguments of Horsthemke and Lefever\cite{Horsthemke2006} in the stationary limit $t\rightarrow\infty$ the term $\frac{\mathrm{d}}{\mathrm{d}t}\overline{\ln|\sin\phi|}$ vanishes and we end up with $\Lambda=\lim\limits_{t\rightarrow\infty}\frac{\mathrm{d}}{\mathrm{d}t}\overline{\rho}=\langle\cot\phi\rangle$.

\section{}

If expressing $z=\cot\phi$ by $v=\cos\phi$, one finds two branches of the solution:
\begin{equation*}
 \cot\phi=\begin{cases}
           \frac{\cos\phi}{\sqrt{1-\cos^2\phi}},\text{ for }\phi\in[0,\pi]\\
           -\frac{\cos\phi}{\sqrt{1-\cos^2\phi}},\text{ for }\phi\in[\pi,2\pi]\quad.
          \end{cases}
\end{equation*}
We take this to consideration in the following way (For simplicity, we only treat the friction-less case $\alpha=0$. However, the general case follows in a similar way): Using $\dot z=-1-z^2-\xi(t)$ to find $\dot \phi$, one ends up with $\dot \phi=1+\xi(t) \sin^2\phi>0,\text{as long as }\xi,-k\xi>-1$.\\
Consequently
\begin{equation*}
 \dot v=\begin{cases}
           -\sqrt{1-v^2}-(1-v^2)^{3/2}\xi(t)<0,\text{ for }\phi\in[0,\pi]\\
           \sqrt{1-v^2}+(1-v^2)^{3/2}\xi(t)>0,\text{ for }\phi\in[\pi,2\pi]\quad.
          \end{cases}
\end{equation*} and $v$ goes from $1$ to $-1$ in the first branch and from $-1$ to $1$ in the second branch. This sense of direction has to be considered for most integrals when calculating $P(v)$. For simplicity, we chose the second branch for the 'natural' direction of integration. However, the other branch would produce symmetric solutions with the same result. As soon as friction is included, the r.h.s. of $\dot\phi$ may get below zero. Although the detailed argumentation gets slightly more complicated in this case, the same arguments hold.
\nocite{*}


\begin{thebibliography}{14}%
\makeatletter
\providecommand \@ifxundefined [1]{%
 \@ifx{#1\undefined}
}%
\providecommand \@ifnum [1]{%
 \ifnum #1\expandafter \@firstoftwo
 \else \expandafter \@secondoftwo
 \fi
}%
\providecommand \@ifx [1]{%
 \ifx #1\expandafter \@firstoftwo
 \else \expandafter \@secondoftwo
 \fi
}%
\providecommand \natexlab [1]{#1}%
\providecommand \enquote  [1]{``#1''}%
\providecommand \bibnamefont  [1]{#1}%
\providecommand \bibfnamefont [1]{#1}%
\providecommand \citenamefont [1]{#1}%
\providecommand \href@noop [0]{\@secondoftwo}%
\providecommand \href [0]{\begingroup \@sanitize@url \@href}%
\providecommand \@href[1]{\@@startlink{#1}\@@href}%
\providecommand \@@href[1]{\endgroup#1\@@endlink}%
\providecommand \@sanitize@url [0]{\catcode `\\12\catcode `\$12\catcode
  `\&12\catcode `\#12\catcode `\^12\catcode `\_12\catcode `\%12\relax}%
\providecommand \@@startlink[1]{}%
\providecommand \@@endlink[0]{}%
\providecommand \url  [0]{\begingroup\@sanitize@url \@url }%
\providecommand \@url [1]{\endgroup\@href {#1}{\urlprefix }}%
\providecommand \urlprefix  [0]{URL }%
\providecommand \Eprint [0]{\href }%
\providecommand \doibase [0]{http://dx.doi.org/}%
\providecommand \selectlanguage [0]{\@gobble}%
\providecommand \bibinfo  [0]{\@secondoftwo}%
\providecommand \bibfield  [0]{\@secondoftwo}%
\providecommand \translation [1]{[#1]}%
\providecommand \BibitemOpen [0]{}%
\providecommand \bibitemStop [0]{}%
\providecommand \bibitemNoStop [0]{.\EOS\space}%
\providecommand \EOS [0]{\spacefactor3000\relax}%
\providecommand \BibitemShut  [1]{\csname bibitem#1\endcsname}%
\let\auto@bib@innerbib\@empty
\bibitem [{\citenamefont {{Mallick}}\ and\ \citenamefont
  {{Marcq}}(2003)}]{MallickMarcq2003}%
  \BibitemOpen
  \bibfield  {author} {\bibinfo {author} {\bibfnamefont {K.}~\bibnamefont
  {{Mallick}}}\ and\ \bibinfo {author} {\bibfnamefont {P.}~\bibnamefont
  {{Marcq}}},\ }\bibfield  {title} {\enquote {\bibinfo {title} {{Stability
  analysis of a noise-induced Hopf bifurcation}},}\ }\href {\doibase
  10.1140/epjb/e2003-00324-y} {\bibfield  {journal} {\bibinfo  {journal}
  {European Physical Journal B}\ }\textbf {\bibinfo {volume} {36}},\ \bibinfo
  {pages} {119--128} (\bibinfo {year} {2003})},\ \Eprint
  {http://arxiv.org/abs/cond-mat/0312360} {arXiv:cond-mat/0312360
  [cond-mat.stat-mech]} \BibitemShut {NoStop}%
\bibitem [{\citenamefont {Mallick}\ and\ \citenamefont
  {Marcq}(2004)}]{MallickMarcq2004}%
  \BibitemOpen
  \bibfield  {author} {\bibinfo {author} {\bibfnamefont {K.}~\bibnamefont
  {Mallick}}\ and\ \bibinfo {author} {\bibfnamefont {P.}~\bibnamefont
  {Marcq}},\ }\bibfield  {title} {\enquote {\bibinfo {title} {On the stochastic
  pendulum with ornstein–uhlenbeck noise},}\ }\href {\doibase
  10.1088/0305-4470/37/17/008} {\bibfield  {journal} {\bibinfo  {journal}
  {Journal of Physics A: Mathematical and General}\ }\textbf {\bibinfo {volume}
  {37}},\ \bibinfo {pages} {4769–4785} (\bibinfo {year} {2004})}\BibitemShut
  {NoStop}%
\bibitem [{\citenamefont {{Mallick}}\ and\ \citenamefont
  {{Peyneau}}(2006)}]{MallickPeyneau2006}%
  \BibitemOpen
  \bibfield  {author} {\bibinfo {author} {\bibfnamefont {K.}~\bibnamefont
  {{Mallick}}}\ and\ \bibinfo {author} {\bibfnamefont {P.-E.}\ \bibnamefont
  {{Peyneau}}},\ }\bibfield  {title} {\enquote {\bibinfo {title} {{Phase
  diagram of the random frequency oscillator: The case of Ornstein Uhlenbeck
  noise}},}\ }\href {\doibase 10.1016/j.physd.2006.07.013} {\bibfield
  {journal} {\bibinfo  {journal} {Physica D Nonlinear Phenomena}\ }\textbf
  {\bibinfo {volume} {221}},\ \bibinfo {pages} {72--83} (\bibinfo {year}
  {2006})},\ \Eprint {http://arxiv.org/abs/cond-mat/0608049}
  {arXiv:cond-mat/0608049 [cond-mat.stat-mech]} \BibitemShut {NoStop}%
\bibitem [{\citenamefont {{Howard}}, \citenamefont {{Hor{\'{a}}nyi}},\ and\
  \citenamefont {{Stewart}}(1999)}]{Howard1999}%
  \BibitemOpen
  \bibfield  {author} {\bibinfo {author} {\bibfnamefont {J.~E.}\ \bibnamefont
  {{Howard}}}, \bibinfo {author} {\bibfnamefont {M.}~\bibnamefont
  {{Hor{\'{a}}nyi}}}, \ and\ \bibinfo {author} {\bibfnamefont {G.~R.}\
  \bibnamefont {{Stewart}}},\ }\bibfield  {title} {\enquote {\bibinfo {title}
  {{Global Dynamics of Charged Dust Particles in Planetary Magnetospheres}},}\
  }\href {\doibase 10.1103/PhysRevLett.83.3993} {\bibfield  {journal} {\bibinfo
   {journal} {\prl}\ }\textbf {\bibinfo {volume} {83}},\ \bibinfo {pages}
  {3993--3996} (\bibinfo {year} {1999})}\BibitemShut {NoStop}%
\bibitem [{\citenamefont {{Hsu}}\ \emph {et~al.}(2011)\citenamefont {{Hsu}},
  \citenamefont {{Postberg}}, \citenamefont {{Kempf}}, \citenamefont
  {{Trieloff}}, \citenamefont {{Burton}}, \citenamefont {{Roy}}, \citenamefont
  {{Moragas-Klostermeyer}},\ and\ \citenamefont {{Srama}}}]{Hsu2011}%
  \BibitemOpen
  \bibfield  {author} {\bibinfo {author} {\bibfnamefont {H.~W.}\ \bibnamefont
  {{Hsu}}}, \bibinfo {author} {\bibfnamefont {F.}~\bibnamefont {{Postberg}}},
  \bibinfo {author} {\bibfnamefont {S.}~\bibnamefont {{Kempf}}}, \bibinfo
  {author} {\bibfnamefont {M.}~\bibnamefont {{Trieloff}}}, \bibinfo {author}
  {\bibfnamefont {M.}~\bibnamefont {{Burton}}}, \bibinfo {author}
  {\bibfnamefont {M.}~\bibnamefont {{Roy}}}, \bibinfo {author} {\bibfnamefont
  {G.}~\bibnamefont {{Moragas-Klostermeyer}}}, \ and\ \bibinfo {author}
  {\bibfnamefont {R.}~\bibnamefont {{Srama}}},\ }\bibfield  {title} {\enquote
  {\bibinfo {title} {{Stream particles as the probe of the
  dust-plasma-magnetosphere interaction at Saturn}},}\ }\href {\doibase
  10.1029/2011JA016488} {\bibfield  {journal} {\bibinfo  {journal} {Journal of
  Geophysical Research (Space Physics)}\ }\textbf {\bibinfo {volume} {116}},\
  \bibinfo {eid} {A09215} (\bibinfo {year} {2011})}\BibitemShut {NoStop}%
\bibitem [{\citenamefont {Hsu}\ \emph {et~al.}(2018)\citenamefont {Hsu},
  \citenamefont {Schmidt}, \citenamefont {Kempf}, \citenamefont {Postberg},
  \citenamefont {Moragas-Klostermeyer}, \citenamefont {Sei{\ss}}, \citenamefont
  {Hoffmann}, \citenamefont {Burton}, \citenamefont {Ye}, \citenamefont
  {Kurth}, \citenamefont {Hor{\'{a}}nyi}, \citenamefont {Khawaja},
  \citenamefont {Spahn}, \citenamefont {Schirdewahn}, \citenamefont
  {O{\textquoteright}Donoghue}, \citenamefont {Moore}, \citenamefont {Cuzzi},
  \citenamefont {Jones},\ and\ \citenamefont {Srama}}]{Hsu2018}%
  \BibitemOpen
  \bibfield  {author} {\bibinfo {author} {\bibfnamefont {H.-W.}\ \bibnamefont
  {Hsu}}, \bibinfo {author} {\bibfnamefont {J.}~\bibnamefont {Schmidt}},
  \bibinfo {author} {\bibfnamefont {S.}~\bibnamefont {Kempf}}, \bibinfo
  {author} {\bibfnamefont {F.}~\bibnamefont {Postberg}}, \bibinfo {author}
  {\bibfnamefont {G.}~\bibnamefont {Moragas-Klostermeyer}}, \bibinfo {author}
  {\bibfnamefont {M.}~\bibnamefont {Sei{\ss}}}, \bibinfo {author}
  {\bibfnamefont {H.}~\bibnamefont {Hoffmann}}, \bibinfo {author}
  {\bibfnamefont {M.}~\bibnamefont {Burton}}, \bibinfo {author} {\bibfnamefont
  {S.}~\bibnamefont {Ye}}, \bibinfo {author} {\bibfnamefont {W.~S.}\
  \bibnamefont {Kurth}}, \bibinfo {author} {\bibfnamefont {M.}~\bibnamefont
  {Hor{\'{a}}nyi}}, \bibinfo {author} {\bibfnamefont {N.}~\bibnamefont
  {Khawaja}}, \bibinfo {author} {\bibfnamefont {F.}~\bibnamefont {Spahn}},
  \bibinfo {author} {\bibfnamefont {D.}~\bibnamefont {Schirdewahn}}, \bibinfo
  {author} {\bibfnamefont {J.}~\bibnamefont {O{\textquoteright}Donoghue}},
  \bibinfo {author} {\bibfnamefont {L.}~\bibnamefont {Moore}}, \bibinfo
  {author} {\bibfnamefont {J.}~\bibnamefont {Cuzzi}}, \bibinfo {author}
  {\bibfnamefont {G.~H.}\ \bibnamefont {Jones}}, \ and\ \bibinfo {author}
  {\bibfnamefont {R.}~\bibnamefont {Srama}},\ }\bibfield  {title} {\enquote
  {\bibinfo {title} {In situ collection of dust grains falling from
  saturn{\textquoteright}s rings into its atmosphere},}\ }\href {\doibase
  10.1126/science.aat3185} {\bibfield  {journal} {\bibinfo  {journal}
  {Science}\ }\textbf {\bibinfo {volume} {362}} (\bibinfo {year} {2018}),\
  10.1126/science.aat3185},\ \Eprint
  {http://arxiv.org/abs/https://science.sciencemag.org/content/362/6410/eaat3185.full.pdf}
  {https://science.sciencemag.org/content/362/6410/eaat3185.full.pdf}
  \BibitemShut {NoStop}%
\bibitem [{\citenamefont {{Casetti}}, \citenamefont {{Livi}},\ and\
  \citenamefont {{Pettini}}(1995)}]{Casetti1995}%
  \BibitemOpen
  \bibfield  {author} {\bibinfo {author} {\bibfnamefont {L.}~\bibnamefont
  {{Casetti}}}, \bibinfo {author} {\bibfnamefont {R.}~\bibnamefont {{Livi}}}, \
  and\ \bibinfo {author} {\bibfnamefont {M.}~\bibnamefont {{Pettini}}},\
  }\bibfield  {title} {\enquote {\bibinfo {title} {{Gaussian Model for Chaotic
  Instability of Hamiltonian Flows}},}\ }\href {\doibase
  10.1103/PhysRevLett.74.375} {\bibfield  {journal} {\bibinfo  {journal}
  {\prl}\ }\textbf {\bibinfo {volume} {74}},\ \bibinfo {pages} {375--378}
  (\bibinfo {year} {1995})}\BibitemShut {NoStop}%
\bibitem [{\citenamefont {{Zillmer}}\ and\ \citenamefont
  {{Pikovsky}}(2003)}]{Zillmer2003}%
  \BibitemOpen
  \bibfield  {author} {\bibinfo {author} {\bibfnamefont {R.}~\bibnamefont
  {{Zillmer}}}\ and\ \bibinfo {author} {\bibfnamefont {A.}~\bibnamefont
  {{Pikovsky}}},\ }\bibfield  {title} {\enquote {\bibinfo {title}
  {{Multiscaling of noise-induced parametric instability}},}\ }\href {\doibase
  10.1103/PhysRevE.67.061117} {\bibfield  {journal} {\bibinfo  {journal}
  {\pre}\ }\textbf {\bibinfo {volume} {67}},\ \bibinfo {eid} {061117} (\bibinfo
  {year} {2003})}\BibitemShut {NoStop}%
\bibitem [{\citenamefont {Bourret}, \citenamefont {Frisch},\ and\ \citenamefont
  {Pouquet}(1973)}]{Bourret1973}%
  \BibitemOpen
  \bibfield  {author} {\bibinfo {author} {\bibfnamefont {R.}~\bibnamefont
  {Bourret}}, \bibinfo {author} {\bibfnamefont {U.}~\bibnamefont {Frisch}}, \
  and\ \bibinfo {author} {\bibfnamefont {A.}~\bibnamefont {Pouquet}},\
  }\bibfield  {title} {\enquote {\bibinfo {title} {Brownian motion of harmonic
  oscillator with stochastic frequency},}\ }\href {\doibase
  https://doi.org/10.1016/0031-8914(73)90347-9} {\bibfield  {journal} {\bibinfo
   {journal} {Physica}\ }\textbf {\bibinfo {volume} {65}},\ \bibinfo {pages}
  {303 -- 320} (\bibinfo {year} {1973})}\BibitemShut {NoStop}%
\bibitem [{Note1()}]{Note1}%
  \BibitemOpen
  \bibinfo {note} {If the assumption of Eq. \ref {eq:zero-average} is not
  fulfilled, the random process can be transformed to $\xi (t)\rightarrow
  \protect \frac {\xi (t)-\protect \langle \xi (t)\protect \rangle }{1+\protect
  \langle \xi (t)\protect \rangle }$, $\xi _{1,2}\rightarrow \protect \frac
  {\xi _{1,2}-\protect \langle \xi (t)\protect \rangle }{1+\protect \langle \xi
  (t)\protect \rangle }$ and a renormalization of the time $t\rightarrow
  t\protect \sqrt {1+\protect \langle \xi (t)\protect \rangle }$ leads again to
  the equation of motion (Eq. \ref {eq:dichotomicosci}) and $\protect \langle
  \xi (t)\protect \rangle =0$. This implies, that one value $\xi _{1,2}$ is
  negative, the other is positive.}\BibitemShut {Stop}%
\bibitem [{\citenamefont {{Gardiner}}(2009)}]{Gardiner2009}%
  \BibitemOpen
  \bibfield  {author} {\bibinfo {author} {\bibfnamefont {C.}~\bibnamefont
  {{Gardiner}}},\ }\href@noop {} {\emph {\bibinfo {title} {Stochastic methods.
  A handbook for the natural and social sciences. 4th revised and augmented
  ed.}}},\ \bibinfo {edition} {4th}\ ed.\ (\bibinfo  {publisher} {Berlin:
  Springer},\ \bibinfo {year} {2009})\ pp.\ \bibinfo {pages} {xvii +
  447}\BibitemShut {NoStop}%
\bibitem [{\citenamefont {Horsthemke}\ and\ \citenamefont
  {Lefever}(2006)}]{Horsthemke2006}%
  \BibitemOpen
  \bibfield  {author} {\bibinfo {author} {\bibfnamefont {W.}~\bibnamefont
  {Horsthemke}}\ and\ \bibinfo {author} {\bibfnamefont {R.}~\bibnamefont
  {Lefever}},\ }\href {https://books.google.de/books?id=OHDafY2VTQAC} {\emph
  {\bibinfo {title} {Noise-Induced Transitions: Theory and Applications in
  Physics, Chemistry, and Biology}}},\ Springer Series in Synergetics\
  (\bibinfo  {publisher} {Springer Berlin Heidelberg},\ \bibinfo {year}
  {2006})\BibitemShut {NoStop}%
\bibitem [{\citenamefont {Anishchenko}\ \emph {et~al.}(2003)\citenamefont
  {Anishchenko}, \citenamefont {Astakhov}, \citenamefont {Neiman},
  \citenamefont {Vadivasova},\ and\ \citenamefont
  {Schimansky-Geier}}]{Anishchenko2003}%
  \BibitemOpen
  \bibfield  {author} {\bibinfo {author} {\bibfnamefont {V.}~\bibnamefont
  {Anishchenko}}, \bibinfo {author} {\bibfnamefont {V.}~\bibnamefont
  {Astakhov}}, \bibinfo {author} {\bibfnamefont {A.}~\bibnamefont {Neiman}},
  \bibinfo {author} {\bibfnamefont {T.}~\bibnamefont {Vadivasova}}, \ and\
  \bibinfo {author} {\bibfnamefont {L.}~\bibnamefont {Schimansky-Geier}},\
  }\href {https://books.google.de/books?id=pXUWMJ70kVUC} {\emph {\bibinfo
  {title} {Nonlinear Dynamics of Chaotic and Stochastic Systems: Tutorial and
  Modern Developments}}},\ Springer Series in Synergetics\ (\bibinfo
  {publisher} {Springer Berlin Heidelberg},\ \bibinfo {year}
  {2003})\BibitemShut {NoStop}%
\bibitem [{\citenamefont {{L{\'{o}}pez-Renteria}}, \citenamefont {{Verduzco}},\
  and\ \citenamefont {{Aguirre-Hern{\'{a}}ndez}}(2015)}]{Lopez2015}%
  \BibitemOpen
  \bibfield  {author} {\bibinfo {author} {\bibfnamefont {J.~A.}\ \bibnamefont
  {{L{\'{o}}pez-Renteria}}}, \bibinfo {author} {\bibfnamefont {F.}~\bibnamefont
  {{Verduzco}}}, \ and\ \bibinfo {author} {\bibfnamefont {B.}~\bibnamefont
  {{Aguirre-Hern{\'{a}}ndez}}},\ }\bibfield  {title} {\enquote {\bibinfo
  {title} {{Control of the Hopf Bifurcation by a Linear Feedback Control}},}\
  }\href {\doibase 10.1142/S0218127415500066} {\bibfield  {journal} {\bibinfo
  {journal} {International Journal of Bifurcation and Chaos}\ }\textbf
  {\bibinfo {volume} {25}},\ \bibinfo {eid} {1550006-289} (\bibinfo {year}
  {2015})}\BibitemShut {NoStop}%
\end{thebibliography}
\end{document}